\documentclass[11pt]{article}
\usepackage{amsmath,amsfonts}
\usepackage{braket}
\usepackage{tikz}
\usetikzlibrary{trees,er,snakes,shapes,mindmap}
\def\sloppy{\tolerance=100000\hfuzz=\maxdimen\vfuzz=\maxdimen}
\def \beq  {\begin{equation}}
\def \eeq  {\end{equation}}
\def \beqar {\begin{eqnarray}}
\def \eeqar {\end{eqnarray}}
\def\sqr#1#2{{\vcenter{\vbox{\hrule height.#2pt
\hbox{\vrule width.#2pt height#1pt \kern#1pt
\vrule width.#2pt}\hrule height.#2pt}}}}

\def\S {{\cal S}}
\def\la {{\langle}}
\def\ra {{\rangle}}
\def\vx {{\vec x}}
\def\vy {{\vec y}}

\def\vf {{\varphi}}

\def\dag {{\dagger}}

\def\Tr {{\rm Tr}}

\def\bp {\bar p}

\def\bD {\bar{D}}

\def\bx {\bar{x}}
\def\by {\bar{y}}

\def\vx {{\vec x}}

\def\vy{\vec{y}}

\def\dag {\dagger}
\def\del {\partial}
\def\bdel{\bar{\partial}}

\def\bz {{\bar{z}}}

\def\A {{\cal A}}
\def\C {{\cal C}}
\def\H {{\cal H}}

\def\G {{\cal G}}

\def\E {{\cal E}}

\def\vf {{\varphi}}

\def \C {{\cal C}}

\def \H {{\cal H}}

\begin{document}
%%%%%%%%%%%%%%%%%%%%%%%%%%%%%%%%%%%%%%%
%%%%%%%%%%%%%%%%%%%%%%%%%%%%%%%%%%%%%%%
%\fontfamily{pnb}\fontsize{12pt}{16pt}\selectfont
%\fontfamily{pzc}\fontsize{14pt}{16pt}\selectfont
%\fontfamily{pbk}\fontsize{12pt}{16pt}\selectfont
%\fontfamily{cmr}\fontsize{11pt}{15pt}\selectfont
\fontfamily{bch}\fontsize{12pt}{17pt}\selectfont
%\fontfamily{phv}\fontshape{ro}\fontsize{11pt}{14pt}\selectfont
%\fontfamily{ptm}\fontseries{m}\fontshape{r}\fontsize{12pt}{16pt}\selectfont
%\fontfamily{pnc}\fontseries{m}\fontshape{r}\fontsize{11pt}{15pt}\selectfont
%\fontfamily{ppl}\fontseries{m}\fontshape{r}\fontsize{11pt}{15pt}\selectfont
%\usefont{T1}{phv}{m}{it}
%%%%%%%%%%%%%%%%%%%%%%%%%%%%%%%%%%%%%%%%%%%%%%%
\def \CMP {{Commun. Math. Phys.}}
\def \PRL {{Phys. Rev. Lett.}}
\def \PL {{Phys. Lett.}}
\def \NPBProc {{Nucl. Phys. B (Proc. Suppl.)}}
\def \NP {{Nucl. Phys.}}
\def \RMP {{Rev. Mod. Phys.}}
\def \JGP {{J. Geom. Phys.}}
\def \CQG {{Class. Quant. Grav.}}
\def \MPL {{Mod. Phys. Lett.}}
\def \IJMP {{ Int. J. Mod. Phys.}}
\def \JHEP {{JHEP}}
\def \PR {{Phys. Rev.}}
\def \JMP {{J. Math. Phys.}}
\def \GRG{{Gen. Rel. Grav.}}
%%%%%%%%%%%%%%%%%%%%%%%%%%%%%%%%%%%%%%%
%\title{Towards a Proof of Mass Gap in 3d Yang-Mills Theory}

%\author{V. P. Nair}
%\affiliation{Physics Department, City College of the City University of New York, New York, NY 10031}

%\begin{abstract}

%\end{abstract}
%\maketitle
%%%%%%%%%%%%%%%%%%%%%%%%%%%%%%%%%%%%%%%
%%%%%%%%%%%%%%%%%%%%%%%%%%%%%%%%%%%%%%%
\begin{titlepage}
\null\vspace{-62pt} \pagestyle{empty}
\begin{center}
%\rightline{CCNY-HEP-18/4}
%\rightline{August 2018}
\vspace{1truein} {\Large\bfseries
Towards a Proof of Mass Gap in 3d Yang-Mills Theory}\\
\vskip .15in
{\Large\bfseries ~}\\
%\vskip .1in
%{\Large\bfseries ~}\\
%%%%%%%%%%%%%%%%%%%%%%%%%%%%%%%%%%%%%%%
%%%%%%%%%%%%%%%%%%%%%%%%%%%%%%%%%%%%%%%
\vspace{.2in}
 {\large\sc V.P. Nair}\\
\vskip .2in
{\itshape Physics Department\\
City College of the City University of New York\\
New York, NY 10031}\\
\vskip .1in
E-mail:~
{\fontfamily{cmtt}\fontsize{11pt}{15pt}\selectfont vnair@ccny.cuny.edu}

\vspace{.8in}
%\vspace{1.5in}
%\vspace{0.3in}
%%%%%%%%%%%%%%%%%%%%%%%%%%%%%%%%%%%%%%%
%%%%%%%%%%%%%%%%%%%%%%%%%%%%%%%%%%%%%%%
\centerline{\large\bf Abstract}
\end{center}
We give the outlines of a proof for the existence of a mass gap for the
3d Yang-Mills theory in terms of an inequality on the eigenvalues of the
Laplacian on the gauge-invariant configuration space.

\end{titlepage}
%%%%%%%%%%%%%%%%%%%%%%%%%%%%%%%%%%%%%%%
%%%%%%%%%%%%%%%%%%%%%%%%%%%%%%%%%%%%%%%
\pagestyle{plain} \setcounter{page}{2}
%\section{Introduction}
Understanding Yang-Mills theories has been a persistent theme of research in particle physics for well over five decades now.
While many important features can be considered as, more or less, understood, the existence of a mass gap is still very much an open question,
despite being an issue of seminal value for all nonperturbative
aspects of the theory.
A key expected feature, based on qualitative arguments, for pure Yang-Mills theories with no spontaneous symmetry breaking was the property of confinement meaning that the physical states will have zero charge corresponding to the symmetry group.
A related consequence is that the lowest lying excitation above the vacuum state will have a nonzero mass. This feature of a mass gap is expected
to hold true in three dimensions as well as for the more realistic four-dimensional case. (We may note that the three dimensional (3d)  theory also does carry some modest relevance to reality since, in addition to serving as a
simpler model for theoretical studies of the 4d theory, it is useful for describing the high temperature phase of the 4d theory.)
The mass gap in 3d Yang-Mills theory is the subject of this paper.

We will follow the Hamiltonian approach which has been developed over the last several years \cite{KN1}-\cite{nair-ias}. The key advantage is that, the spatial manifold being two-dimensional, one is able to utilize some known exact results regarding 2d gauge theories. The relevant configuration space $\C = \A /\G_*$ 
is the set of all
2d gauge potentials ($\A$) modulo the set of all gauge transformations
which tend to the identity at spatial infinity ($\G_*$).
It is possible to derive an exact expression for the volume element
for the gauge-orbit space $\C$.
 It is expected to play a crucial role in generating a mass gap and, in fact, heuristic arguments towards this were given many years ago \cite{{KN1},{KN2}}. The Hamiltonian is the sum of a kinetic energy operator $T$ and a potential energy $V$. The operator $T$ can be taken as the
Laplacian on the (infinite-dimensional) gauge-orbit space $\C$
while $V$ is the integral of the square of the magnetic field. Both operators need regularization to make them well-defined. While the Laplacian on a compact finite-dimensional manifold has a gap in its spectrum, an extension to the present problem (as suggested by Feynman \cite{feynman}) is not obtained since, strictly speaking, $\C$ is not compact.
Since the Lichnerowicz inequality for the spectrum of the Laplacian
involves the Ricci curvature, 
Singer suggested proving the positivity of this curvature on
$\C$ \cite{singer}. This can be shown in a neighborhood of flat connections
on $\C$, but difficulties of regularization prevent a general proof.
More recently, there have been suggestions on using the Bakry-Emery curvature \cite{mondal}.
Expectation values in the theory involve integration over $\C$ with
a weight factor $\vert \Psi_0\vert^2$, where $\Psi_0$ is the vacuum wave function. The Hessian of $\log \vert \Psi_0\vert^2$ added to
the Ricci curvature is the Bakry-Emery curvature and this is what naturally enters the generalization of the Lichnerowicz inequality.
Properties of $\vert \Psi_0\vert^2$ can help towards a proof of the mass gap.
Once again, regularization issues prevent a general proof.
Nevertheless, one can make some nontrivial statements based on the
form of $\vert \Psi_0\vert^2$ given in \cite{KKN2}.

We will use an algebraic approach in this paper. We will derive a set of algebraic relations for various operators. This will be informed by 
the geometric properties of $\C$, in particular, the volume element on $\C$.
One can then obtain an inequality on the spectrum of the Laplacian.
%%%%%%%%%%%%%%%%%%%%%%%%%%%%%%%%%%%%%%%
\vskip .1in\noindent{\underline{Basic framework}}
\vskip .1in
%%%%%%%%%%%%%%%%%%%%%%%%%%%%%%%%%%%%%%%
Turning to specifics, we will consider a $G= SU(N)$-gauge theory, the gauge
potentials are of the form
$A_\mu = -i t_a A_\mu^a$, $\mu = 0, 1, 2$,
where $t_a$ are hermitian 
$N \times N$-matrices which form a basis of the Lie algebra of $SU(N)$ with
$[t_a, t_b ] = i f_{abc}\, t_c$, and ${\Tr} (t_a t_b) = {1 \over 2} \delta_{ab}$. 
In the Hamiltonian approach we choose the 
gauge $A_0 = 0$.
Further we will use complex coordinates
$z = x_1 - i x_2$, $\bz = x_1 +i x_2$ for the spatial manifold
and complex components $A_{z} = {1 \over 2} (A_1 +i A_2), ~~ 
A_{\bar{z}} =
{1 \over 2} (A_1 -i A_2) = - (A_z)^{\dagger}$ for
the spatial components of the potential.
These complex components can be parametrized in terms of a
matrix $M$ as
\beq
A_z = -\partial_{z} M M^{-1},\hskip .3in A_{\bar{z}} = M^{\dagger -1} \partial_
{\bar{z}} M^{\dagger}
\label{gap1}
\eeq
The matrix-valued field $M$ is an element of the complexification of the group $G$, which is $SL(N, \mathbb{C})$ for $G = SU(N)$.

The Hamiltonian is given by $\H = T + V$, where
\beq
T={e^2\over 2} \int d^2 x ~ E_i^a E_i^a, \hskip .3in
V= {1\over 2e^2} \int d^2x~ B^a B^a 
\label{gap2}
\eeq
Here $E^a_i = \del_0 A^a_i$ is the electric field and
$B^a={1 \over 2} \epsilon_{jk}(\partial_j A_k^a - \partial_k A_j^a +f^{abc}
 A_j^b A_k^c )$ is the magnetic field.
 The parameter $e$ is the coupling constant, $e^2$ has the dimension
 of mass.
 We will not go through the details of the canonical quantization of the theory as it has been worked out in \cite{{KN1},{KKN1}}. The key point is that the 
 two components of the electric field are given by
 \beqar
 E_{z\,k}  &=& ~{i\over 2} {\E}^\dagger_{ak}\int _{x'} \bar{G} (x, x') \bar{p}_a (x') \nonumber\\
E_{\bar{z}\,k}  &=& -{i\over 2} {\E}_{ka}\int _{x'} G(x, x') p_a(x')
\label{gap3}
\eeqar
where $G(x, x')$ and ${\bar G}(x, x')$ are Green's functions for
$\del_z$ and $\bdel_\bz$ respectively, obeying
\beq
 \bar{\partial}_x \bar{G} (x, x') ~= \partial _x G(x, x') ~= \delta ^{(2)} (x-x')
\label{gap4}
\eeq
If needed, explicit formulae for these can be taken as
\beq
 \bar{G} (x,x') = {1 \over {\pi (z-z')}}, \hskip .3in
G(x,x') = {1 \over {\pi (\bar{z} - \bar{z}')}}
\label{gap5}
\eeq
Also, in (\ref{gap3}), we have defined
\beq
{\E}_{ka} = 2 ~\Tr (t_k M t_a M^{-1}),\hskip .3in {\E}^{\dagger} _{ak} = 2 ~\Tr (t_a M^{\dagger}
t_k M^{\dagger -1}) 
\label{gap6}
\eeq
These are the $SL(N, \mathbb{C})$ matrices $M$ and $M^\dagger$ in the adjoint representation.
Further, $p_a$ and $\bp_a$ are translation operators on $M$ and
$M^\dagger$ respectively. The canonical commutation rules in terms of these operators are
\begin{align}
[p_a (x), M (y) ] & = -i M (x)\, t_a ~ \delta ^{(2)} (x-y) \cr
[p_a (x), p_b (y) ] & = f_{abc} p_c(x)~ \delta ^{(2)} (x-y) \cr
[\bar{p}_a (x), M^{\dagger} (y) ] & = -i t_a\, M^{\dagger} (x)~ \delta ^{(2)} 
(x-y) \cr
[\bar{p}_a (x), \bar{p}_b (y) ] & = -f_{abc} \bar{p}_c (x) ~ \delta ^{(2)} (x-y) 
\label{gap7}\\
[M(x), M (y) ] & = ~ [M^{\dagger} (x), M^{\dagger} (y)]~=~ [M(x), M^{\dagger} 
(y)] = 0 \cr
[p_a (x), M^{\dagger}] & =~ [\bar{p}_a (x), M(y)]~=~ [p_a (x), \bar{p}_b (y)] 
 =0 \nonumber
\end{align}

Gauge transformations act on $M$ and $M^\dagger$ as
$M \rightarrow M^g = g \, M$, $M^\dagger \rightarrow (M^\dagger)^g
= M^\dagger\, g^\dagger$, where $g \in G$ is an element of the gauge group.
The gauge-invariant variables are given by $H = M^\dagger M \in G^{\mathbb{C}}/G = SL(N, \mathbb{C})/SU(N)$.
The generator of gauge transformations, i.e., the Gauss law operator,
is given by
\beq
G(\theta) = i \int \theta _k ~ ({\E}_{ka} p_a - {\E}^{\dagger} _{ak} \bar{p}_a) 
\label{gap8}
\eeq
for $g \approx 1 + i \theta^a t_a$. It is easy to verify that
$G(\theta)$ commutes with $p_a$, $\bp_a$ and $H$.
The matrix-valued function $H$ may be taken as coordinatizing the gauge-orbit space
$\C$.

The wave functions for physical states should be invariant under the action of
$G(\theta)$. Therefore they can be taken to be functions of $H$.
On such functions, Gauss law is equivalent to the condition
$\bp_a = K_{ab} p_b$ where $K_{ab} = 2 \, \Tr (t_a H t_b H^{-1})=
\E^\dagger_{ak} \E_{kb}$. (This is the relation $H = M^\dagger M$ in the adjoint representation
of $SL(N, \mathbb{C}) )$.
The action of $p_a$, $\bp_a$ on $H$
can also be simplified as
\begin{align}
[p_a (x), H (y) ] & = -i H (x)\, t_a ~ \delta ^{(2)} (x-y) \cr
[\bar{p}_a (x), H(y) ] & = -i t_a H(x)~ \delta ^{(2)} (x-y) 
\label{gap9}
\end{align}
When acting on functions of 
$H$, $p_a$ and ${\bar p}_a$ can be represented as the functional
differential operators
\begin{align}
p_a(x) &= -i r^{-1}_{ab}(x) {\delta \over {\delta \vf^b (x)}}\cr
{\bar p}_a(x) &= -i r^{*-1}_{ab}(x) {\delta \over {\delta \vf^b (x)}}~=
K_{ab}(x)p_b(x)
\label{gap10}
\end{align}
where we parametrize $H$ in terms of the real fields $\vf^a$ and 
\beq
H^{-1} \delta H~= \delta \vf^a r_{ab}~ t^b
\label{gap11}
\eeq
This equation defines $r_{ab}$.

It is straightforward to work out the volume element for the gauge-orbit space
$\C$. Starting with the metric $ds^2 = \int d^2x\, \delta A^a_i \delta A^a_i$
on the space of gauge potentials $\A$, we find \cite{GK}
\beq
d\mu (\C) = d\mu (H)\, \det(-\bD_\bz D_z ) =
\sigma \, d\mu (H) \, e^{2 c_{\rm A} \S(H)} 
\label{gap12}
\eeq
Here $\sigma = [\det (-\bdel \del) /\int d^2x]^{{\rm dim}G}$
is a constant which will not be relevant for us. Further
\beq
d\mu(H) = \prod_x \det r \, [d\vf]
\label{gap13}
\eeq
is the product of Haar measures for $SL(N, \mathbb{C})/SU(N)$.
$\S(H)$ is the Wess-Zumino-Witten (WZW) action for $H$ given by
\beqar
{\S} (H) &=& {1 \over {2 \pi}} \int \Tr (\partial H \bar{\partial} H^{-1})
\label{gap14}\\
&&+{i \over {12 \pi}} \int \epsilon ^{\mu \nu \alpha} \Tr ( H^{-1} \partial _{\mu}
H H^{-1} \partial _{\nu}H H^{-1} \partial _{\alpha}H)
\nonumber
\eeqar
The parameter $c_{\rm A}$ in (\ref{gap12}) is the quadratic Casimir invariant for the adjoint representation of $G$, $c_{\rm A} \delta_{ab} = f_{apq} f_{bpq} = N
\delta_{ab}$ 
for $SU(N)$. The inner product for the wave functions is thus given by
\beq
\braket{1|2} = \int d\mu (H) \, e^{2 c_{\rm A} \S(H)} \,
\Psi_1^* (H)\, \Psi_2(H)
\label{gap15}
\eeq
With this inner product and equations (\ref{gap3}), a general matrix element
of $T$ can be written as
\beqar
\la 1|T|2\ra &=&
{e^2\over 4}\int d\mu (H)~ \Psi_1^* \Bigl[
(\bar{G}\bar{p}_a) ( K_{ab}e^{2 c_A \S} G p_b )\nonumber\\
&&\hskip .2in + (G p _a) K^{T}_
{ab} e^{2 c_A \S}(\bar{G} \bar{p}_b) \Bigr]  \Psi_2
\label{gap16}
\eeqar
The Green's functions $G$ and $\bar{G}$ have to be replaced
by regularized versions to get a well-defined expression for $T$
as discussed in \cite{KKN1}.
Regularization is important if we move $p_a$ and $\bp_a$ to the right end of the operator expression given above since $[ Gp_a, K_{ab}]$ and
$[\bar{G} \bp_a , K_{ab}]$ will lead to singular
expressions. But for many of the commutation rules used below, regularization will not play a crucial role; we will indicate steps where it is crucial to take account of regularization. 

The WZW action in the inner product will be the key factor for the mass gap.
To account for this in a simple algebraic way, we define the wave functions
$\Phi$ by $\Psi = e^{- c_{\rm A} \S} \Phi$, so that the inner product now reads
\beq
\braket{1|2} = \int d\mu (H)\, \Phi_1^* \Phi_2
\label{gap17}
\eeq
The kinetic energy operator, as an operator on $\Phi$'s,  can now be written as
\beq
T  = {e^2 \over 4} \int ( P_a^{\dagger} K_{ab} P_b + Q_a^{\dagger} K^{T}_{ab}
Q_b) 
\label{gap18}
\eeq
where
\begin{align}
&P_a  = \int G (p_a - c_A~ p_a \S), \hskip .3in
P_a^{\dagger}  = \int \bar{G} (\bar{p}_a + c_A ~\bar{p}_a \S) \cr
&Q_a  = \int \bar{G} (\bar{p}_a - c_A~ \bar{p}_a \S), \hskip .3in
Q_a^{\dagger}  = \int G (p_a + c_A~ p_a \S)
\label{gap19}
\end{align}
We can also write the explicit expressions
\beq
p_a {\S} = -{i \over {\pi}}  \Tr [ t_a \partial (H^{-1} \bar{\partial} H)],
\hskip .1in
\bar{p}_a {\S} = -{i \over {\pi}}\Tr [ t_a \bar{\partial} (\partial H H^{-1})] 
\label{gap20}
\eeq
The second term in (\ref{gap18}) can be rewritten as follows.
Using $p_a = K_{ca} \bp_c$, we get
\beqar
&&\int_z [G(x,z) p_a(z) , K_{ba}(x) ]\nonumber\\
&& =
\int_z G(x,z) K_{ca} (z) [\bp_c(z), K_{ba}(x)]\nonumber\\
&&=- \int_z G(x,z) f_{cbl} K_{ca} (z) K^T_{al} (x) \Delta (x,z)\nonumber\\
&&=-\int_z G(x,z) f_{cbl}  \Bigl( \delta_{cl} + (\bz - \bx) (\bdel K K^T)_{cl}
\nonumber\\
&&\hskip .2in+ \cdots\Bigr) \Delta (x,z)\nonumber\\
&&={i \over \pi} c_{\rm A} (\bdel H H^{-1})_b
\label{gap21}
\eeqar
Here $\Delta (x,z)$ is a regularized version of $\delta^{(2)}(x-z)$. It can be
taken to be a narrow Gaussian in $\vert x-z\vert$ and we can expand
$K_{ca} (z)$ around $x$. 
The limit $\Delta (x,z) \rightarrow \delta^{(2)}(x-z)$ is taken at the end.
The terms indicated by the ellipsis in (\ref{gap21}) vanish in this limit.
Using this relation, and a similar one for $P_a^\dagger K_{ab}$,
it is easy to see that
\beq
Q^\dagger_a K^T_{ab} = K_{ba} P_a, \hskip .3in
P^\dagger_a K_{ab} = K_{ab} Q_a
\label{gap22}
\eeq
Thus we can rewrite $T$ as
\beq
T = {e^2 \over 2} \int P_a^\dagger K_{ab} P_b
= {e^2 \over 2} \int W_a^\dagger P_a
= {e^2\over 2} \int P_a^\dagger W_a
\label{gap23}
\eeq
where $W_a = K_{ab} P_b$, $W_a^\dagger = P_b^\dagger K_{ba}$.
The first expression for $T$ shows that it is a positive semidefinite operator.
Thus its nonzero eigenvalues will be positive, the zero mode will correspond
to the ground (vacuum) state.
The commutation rules for $P_a$, $P_a^\dagger$ are
\beqar
[P_a(x), P_b(y)] &=& -f_{abc} G(x,y) \bigl( P_c(x)- P_c(y) \bigr) \cr
{}[P_a^\dag (x), P_b^\dag (y) ] &=& f_{abc} {\bar G}(x,y) \bigl(P_c^\dag (x)- P_c^\dag (y)
\bigr) \cr
{}[P_a(x), P_b^\dag (y)]& =& {c_A\over \pi} K_{ba}(x) \delta^{(2)} (x-y)
\label{gap24}
\eeqar
We also get
\beq
[W_a^\dagger (x), W_b^\dagger (y) ] = [W_a (x), W_b (y) ] = 0
\label{gap25}
\eeq
There are other relations such as $[W_a^\dagger (x), W_b (y)]$;
we do not display them here as they will not be needed for the argument given below. 
%%%%%%%%%%%%%%%%%%%%%%%%%%%%%%%%%%%%%%%
\vskip .1in\noindent{\underline{Argument for mass gap}}
\vskip .1in
%%%%%%%%%%%%%%%%%%%%%%%%%%%%%%%%%%%%%%%
With this preparation, the argument for the mass gap is straightforward.
Basically, we will need the relations (\ref{gap23})-(\ref{gap25}).
We now consider the following sequence of equations:
\beqar
&&\left({e^2 \over 2}\right)^2 \int_{x,y} W^\dagger_b (y) P^\dagger_a(x)
P_b(y) W_a (x) \nonumber\\
&&=\left({e^2 \over 2}\right)^2 \int_{y}
W^\dagger_b (y) P_b(y) \int_x P^\dagger_a(x) W_a (x) \nonumber\\
&&+ \left({e^2 \over 2}\right)^2 \int_{x,y}
W^\dagger_b (y)\, [P^\dagger_a(x), P_b(y)]\, W_a (x)\nonumber\\
&&= T^2 - {c_{\rm A} \over \pi} \left({e^2 \over 2}\right)^2
\int_x W^\dagger_b (x) K_{ab}(x) W_a (x)\nonumber\\
&&= T^2 - m {e^2 \over 2} \int P_a^\dagger W_a\nonumber\\
&&= T^2 - m T
\label{gap26}
\eeqar
where $m = e^2 c_{\rm A} /(2\pi)$.
For the second equality we used (\ref{gap23}) and the last
of the commutation rules (\ref{gap24}).
We now consider another sequence of relations:
\beqar
&&\int W^\dagger_c (x) P^\dagger_d (y) K_{db}(y) K_{ac}(x) 
P_b (y) W_a(x)\nonumber\\
&&=\int W^\dagger_c (x) W^\dagger_b (y) K_{ac}(x) 
P_b (y) W_a(x)\nonumber\\
&&=\int W^\dagger_b (y) W^\dagger_c (x)  K_{ac}(x) 
P_b (y) W_a(x)\nonumber\\
&&=\int W^\dagger_b (y) P^\dagger_l (x) K_{lc}(x) K_{ac}(x) 
P_b (y) W_a(x)\nonumber\\
&&=\int W^\dagger_b (y) P^\dagger_a (x) 
P_b (y) W_a(x)
\label{gap27}
\eeqar
For the second equality we use the fact that $W^\dagger$'s commute
as in (\ref{gap25}). The definition of $W^\dagger$ then leads to the final expression.
Notice that the last expression is proportional to the left hand side
of (\ref{gap26}). Combining these two equations, we get
\beqar
&&T^2 - m T\nonumber\\
&=& \left({e^2 \over 2}\right)^2  \int [B_{dc} (y,x)]^\dagger
\Big\{ K_{db}(y) K_{ac}(x) \Big\} B_{ba}(y,x)
\nonumber\\
&&B_{ba} (y,x) = P_b (y) W_a (x)\label{gap28}
\eeqar
The quantity in the curly brackets is a positive operator
since $K_{ab} = \E^\dagger_{ak} \E_{kb}$.
Therefore the integral on the right hand side is greater than or equal to
zero. If $\ket{0}$ denotes the vacuum state for $T$, excited states are of the form $f \ket{0}$ for some composite operator $f$. We then get
\beqar
&&\bra{0} f^\dagger (T^2 - m T) f \ket{0} \nonumber\\ 
&&=\left({e^2 \over 2}\right)^2\!\!\!  \int\! \bra{0} f^\dagger\,[B_{dc} (y,x)]^\dagger
\Big\{ K_{db}(y) K_{ac}(x) \Big\} \nonumber\\
&&\hskip .7in  B_{ba}(y,x) \, f \ket{0} \nonumber\\
&&\geq 0
\label{gap29}
\eeqar
For an eigenstate of $T$ with eigenvalue $\lambda$, we have
$T \, f \ket{0} = \lambda \, f \ket{0}$. This inequality then becomes
\beq
\lambda ( \lambda - m) \geq 0
\label{gap30}
\eeq
$\lambda = 0$ will saturate this inequality and will correspond to the ground state of $T$, with
$T \ket{0} = 0$, $P_a \ket{0} = 0$. This is equivalent to
$\Psi_0 =1$.\footnote{The normalization integral for this state is 
$\int d\mu (H) e^{ 2 c_{\rm A} \S(H)}$, which is the partition function for a $G^{\mathbb{C}}/G$ WZW model. This is rendered finite by a suitable cutoff on the number of modes. For any wave function in field theory, it is standard
to have such a regularization to define the normalization.}
For all excited states with $\lambda > 0$, we get
\beq
\lambda \geq m
\label{gap31}
\eeq

The Hamiltonian should include the potential energy $V$ as well.
The short distance singularity in the operator product
$B^a(x) B^a (x)$ has to be regularized and subtracted to define
a renormalized operator $V$.
Assuming this can be done maintaining positivity
of $V$, we have $\H \geq T$ and $\lambda \geq m$ for eigenvalues of the full Hamiltonian.

We have shown elsewhere that $T J_a = m J_a$,
where $J_a = (c_{\rm A}/\pi) \del H H^{-1}$. This would seemingly
give a state which saturates
the bound (\ref{gap31}). However $J_a$ is not the wave function
for an acceptable physical state. This is because the parametrization
(\ref{gap1}) does not define $M$, $M^\dagger$ uniquely; $M \bar{V}(\bz )$,
$V(z) M^\dagger$, where $V(z)$ (resp. ${\bar V}$) is holomorphic
(resp. antiholomorphic) will give the same potentials $A_i$.
Therefore we must require invariance of $\Psi$ under the
transformation $H \rightarrow V H {\bar V}$. Minimally we need at least two
powers of $J$ with a wave function of the form \cite{KKN1}
\beq
\Psi _2 = \int _{x,y} f(\vx,\vy) \bigl[ \bdel J_a (\vx) \bigl( K(x,\by) K^{-1} (y,
\by) \bigr) _{ab} \bdel J_b (\vy)  \bigr] 
\label{gap32}
\eeq
The energy for this will be significantly higher than
the bound in (\ref{gap31}).
We expect that there could be another more stringent inequality
where the lower bound is 
higher than $m$.
%%%%%%%%%%%%%%%%%%%%%%%%%%%%%%%%%%%%%%%
\vskip .1in\noindent{\underline{Comparison with an intuitive argument}}
\vskip .1in
%%%%%%%%%%%%%%%%%%%%%%%%%%%%%%%%%%%%%%%
It is also useful to make a comparison with the intuitive or heuristic argument for the mass gap given in
\cite{{KN2},{AN}}.
From the uncertainty principle we can write
\beq
\la {\cal H}\ra = {1\over 2} \left[e^2 \Delta E^2 + {\Delta B^2 \over e^2}\right]
= {1\over 2} \left[{e^2 k^2 \over \Delta B^2}+ {\Delta B^2 \over e^2}\right]
\label{gap33}
\eeq
Here we consider modes of $E$, $B$ fields corresponding to a momentum value $k$. Usually we look for the low lying modes which
can be obtained by minimizing $\la \H\ra$
with respect to $\Delta B^2$. This gives $\Delta B^2 \sim e^2 k$
and hence $\la {\cal H}\ra \sim k$. (This would be the photon in the Abelian theory.) 
However, in our case, the integration
measure in the inner product
(\ref{gap15}) controls the dispersion in $B$.
This is because, for low values of $k$, it becomes a very narrow Gaussian, since
\beq
\S(H) \approx  \left[ -{c_A\over 2\pi} \int B {1\over {k^2}} B +...\right]
\label{gap34}
\eeq
The resulting value of $\Delta B^2 = \pi k^2 /c_A$ leads to
$\la {\cal H}\ra = (e^2 c_A /2\pi) + {\cal O} (k^2)$. 
We see how the integration measure can lead to
a mass gap. The relation $\Delta E^2\sim 1/\Delta B^2$ 
follows from the fact that $E^2$ is a Laplacian on $\A/\G_*$ and so
the potential energy does not play any role
in this argument.
 (If we minimize $\la \H \ra$ balancing between 
$T$ and $V$ (as for the photon), the potential energy is important.)
We should therefore expect that a gap can be obtained just for
the kinetic term $T$. The inequality (\ref{gap31}) may be viewed as
a more precise mathematical realization of this intuitive argument.
By redefining the wave functions as
$\Psi = e^{- c_{\rm A} \S} \Phi$ we have absorbed the effect of
the factor $e^{2 c_{\rm A}\S}$ in the inner product into the operators
$P_a$, $P^\dagger_a$. The effect of the factor $e^{2 c_{\rm A}\S}$
is thus included in the
commutation rules (\ref{gap24}) and the derivation
of the inequality (\ref{gap31}) could then be done
more or less purely algebraically.
%%%%%%%%%%%%%%%%%%%%%%%%%%%%%%%%%%%%%%%
\vskip .1in\noindent{\underline{Relation to older work}}
\vskip .1in
%%%%%%%%%%%%%%%%%%%%%%%%%%%%%%%%%%%%%%%
The basic framework outlined at the beginning has been used to get an approximate solution for the vacuum wave function and the string tension
in \cite{{KKN2},{KKN1}}. In \cite{KNY}, an expansion scheme for 
systematizing corrections to this result was set up in terms of a two-parameter set of theories ${\widetilde{YM}}(m, e^2)$ where $m$ and $e^2$ were treated as independent parameters. The line $m = e^2 c_{\rm A}/(2\pi)$
corresponds to Yang-Mills theory. While this helped to systematize the corrections in powers of $e^2/m$, once we set $m = e^2 c_{\rm A}/(2\pi)$, we do not retain parametric control on the corrections.
The smallness of the corrections found in \cite{KNY} was based on the decreasing value of an integral of the form $\int (d^2k/m^2) (m /\sqrt{k^2 + m^2})^n$
with increasing $n$.
However, we emphasize that the existence of a mass gap based on eigenvalues of $T$, as discussed in this paper, is not sensitive to the issue of an expansion parameter or parametric control over the expansion scheme.

\bigskip

This research was supported in part by the U.S.\ National Science
Foundation grant PHY-2412479.

%%%%%%%%%%%%%%%%%%%%%%%%%%%%%%%%%%%%%%%
%%%%%%%%%%%%%%%%%%%%%%%%%%%%%%%%%%%%%%%

%%%%%%%%%%%%%%%%%%%%%%%%%%%%%%%%%%%%%%%
%%%%%%%%%%%%%%%%%%%%%%%%%%%%%%%%%%%%%%%
%%%%%%%%%%%%%%%%%%%%%%%%%%%%%%%%%%%%%%%
%%%%%%%%%%%%%%%%%%%%%%%%%%%%%%%%%%%%%%%
\end{document}